\documentclass{sigcomm-alternate}

\usepackage{algorithmic}
\usepackage{algorithm}
\usepackage{graphicx}
\usepackage{subfigure}
\usepackage{paralist}

\begin{document}
%

\title{Bidirectional Pipelining \\for Scalable IP Lookup and Packet Classification}
%
%
%
%
%

\numberofauthors{1} 
%
\author{
%
%
\alignauthor
Weirong Jiang, Hoang Le and Viktor K. Prasanna\\
       \affaddr{Ming Hsieh Department of Electrical Engineering}\\
       \affaddr{University of Southern California}\\
       \affaddr{Los Angeles, CA 90089, USA}\\
       \email{\{weirongj, hoangle, prasanna\}@usc.edu}
}

\maketitle
\begin{abstract}
Both IP lookup and packet classification in IP routers can be implemented by some form of tree traversal. SRAM-based Pipelining can improve the throughput dramatically. However, previous pipelining schemes result in unbalanced memory allocation over the pipeline stages. This has been identified as a major challenge for scalable pipelined solutions. This paper proposes a flexible bidirectional linear pipeline architecture based on widely-used dual-port SRAMs. A search tree is partitioned, and then mapped onto pipeline stages by a bidirectional fine-grained mapping scheme. We introduce the notion of \textit{inversion factor} and several heuristics to invert subtrees for memory balancing. Due to its linear structure, the architecture maintains packet input order, and supports non-blocking route updates. Our experiments show that, the architecture can achieve a \textit{perfectly} balanced memory distribution over the pipeline stages, for both trie-based IP lookup and tree-based multi-dimensional packet classification. For IP lookup, it can store a full backbone routing table with 154419 entries using 2MB of memory, and sustain a high throughput of 1.87 billion packets per second (GPPS), i.e. 0.6 Tbps for the minimum size (40 bytes) packets. The throughput can be improved further to be 2.4 Tbps, by employing caching to exploit the Internet traffic locality. 

\end{abstract}




\section{Introduction}
\label{sec:intro}

Modern IP routers need to offer not only IP lookup for packet forwarding, but also a variety of value-added functions, such as security and differentiated services. Most of those functionalities rely on packet classification where the packets are classified into different flows according to some set of pre-defined rules. Packet classification generally refers to the multi-field matching. IP lookup can be seemed as one dimensional packet classification where the destination IP address of a packet is matched against a set of prefixes. 

On the other hand, advances in optical networking technology pose a big challenge on the design of high speed IP routers. Increasing link rates demand that packet processing in IP routers must be performed in hardware. For instance, 40 Gbps links require a throughput of 8 ns per packet, i.e. 125 million packets per second (MPPS), for a minimum size (40 bytes) packet. Such throughput is impossible using existing software-based solutions for either IP lookup \cite{networkmag01:sanchez} or packet classification \cite{networkmag01:gupta}. 
  
Most hardware-based solutions for high speed packet processing in routers fall into two main categories: ternary content addressable memory (TCAM)-based and dynamic/static random access memory (DRAM/SRAM)-based solutions. Although TCAM-based engines can retrieve results in just one clock cycle, their throughput is limited by the relatively low clock rate of TCAMs. TCAMs are expensive and offer little flexibility to adapt to new addressing and routing protocols \cite{infocom08:jiang}. As shown in Table \ref{tb:tcamvssram}, SRAMs offer better scalability than TCAMs with respect to memory access time, density and power consumption. 
  
\begin{table}[htb]
\caption{Comparison of TCAM and SRAM technologies (based on 18 Mbit chip)}
\label{tb:tcamvssram}
\vspace{-0.11in}
\begin{center}
\begin{tabular}{|c|c|c|}
\hline
 & TCAM  & SRAM \\
\hline
\hline
Maximum clock rate (MHz) & 266  \cite{renesas} & 400 \cite{cypress, samsung} \\
\hline
Cell size (\# of transistors/bit) & 16 & 6 \\
\hline
Power consumption (Watts) & 12 $\sim$ 15 \cite{ton06:zheng} & $\approx$ 0.1 \cite{cacti}\\
\hline
\end{tabular}
\end{center}
\end{table}

In DRAM/SRAM-based solutions, both IP lookup and packet classification can be implemented by some form of tree traversal. Each packet traverses a search tree in the memory, and retrieves its matched entry when it arrives at a tree leaf. Such a search process needs multiple memory accesses, which becomes a major drawback of traditional DRAM/SRAM-based solutions. Several researchers have explored pipelining to improve the throughput. A simple pipelining approach is to map each tree level onto a pipeline stage with its own memory and processing logic. One packet can be processed every clock cycle. However, this approach results in unbalanced tree node distribution over the pipeline stages. This has been identified as a dominant issue for pipelined architectures \cite{infocom03:basu}. In an unbalanced pipeline, the ``fattest'' stage, which stores the largest number of tree nodes, becomes a bottleneck. It adversely affects the overall performance of the pipeline in the following aspects. First, more time is needed to access the larger local memory. This leads to a reduction in the global clock rate. 
Second, a fat stage results in many updates, due to the proportional relationship between the number of updates and the number of tree nodes stored in that stage. Particularly during the update process caused by intensive route/rule insertion, the fattest stage may also result in memory overflow. Furthermore, since it is unclear at hardware design time which stage will be the fattest, we need to allocate memory with the maximum size for each stage. Such a kind of over-provisioning results in memory wastage \cite{isca05:baboescu} and excessive power consumption.

To balance the memory distribution across stages, several novel pipeline architectures have been proposed \cite{isca05:baboescu, ancs06:kumar, hoti07:jiang}. However, none of them can achieve a perfectly balanced memory distribution over stages. Furthermore, due to the non-linear structures they employ, most of them must stall subsequent packets during a route update. 

We propose a SRAM-based bidirectional linear pipeline architecture, for both IP lookup and packet classification in IP routers. This paper makes the following contributions:
\begin{itemize}
\item To the best of our knowledge, this work is the first one to achieve a perfectly balanced memory allocation over pipeline stages, for both IP lookup and multi-dimensional packet classification. The memory wastage due to over-provisioning is almost zero.
\item A bidirectional fine-grained mapping scheme is proposed to realize the above goal. We introduce the notion of \textit{inversion factor} and several heuristics to invert the subtrees for memory balancing. 
\item A novel bidirectional linear pipeline architecture is presented to enable the above mapping. It maintains the packet input order and supports non-blocking updates.
\item Our simulation experiments using real-life data demonstrate the SRAM-based pipelined architecture to be a promising solution for next generation IP routers. The proposed architecture can store a full backbone routing table with 154419 entries using 2MB of memory. It can sustain a high throughput of 1.87 billion packets per second (GPPS), i.e. 0.6 Tbps for minimum size (40 bytes) packets.
\end{itemize}

The remainder of this paper is organized as follows. Section \ref{sec:bg} reviews the background and related works. Section \ref{sec:membalance} discusses the memory balancing over pipeline stages and proposes a novel bidirectional fine-grained mapping scheme. Section \ref{sec:arch} proposes a corresponding bidirectional linear pipeline architecture. Section \ref{sec:performance} conducts simulation experiments to evaluate the performance of our approaches. Section \ref{sec:conclusion} concludes the paper.

\section{Background}
\label{sec:bg}

\subsection{IP Lookup and Packet Classification}

\subsubsection{Trie-based IP Lookup} 
\begin{figure}[htb]
\centering
\includegraphics[width=3.0in]{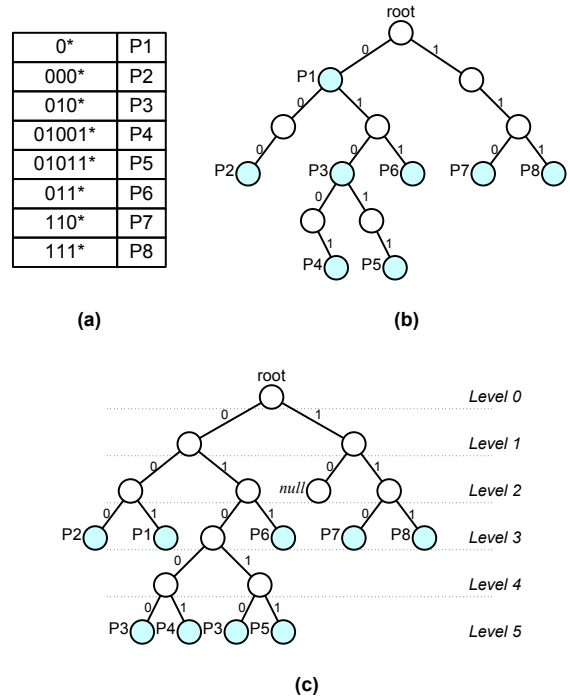}
\caption{(a) Prefix set; (b) Uni-bit trie; (c) Leaf-pushed uni-bit trie.}
\label{fig:trie}
\end{figure}

The nature of IP lookup is longest prefix matching (LPM). The most common data structure in algorithmic solutions for performing LPM is some form of trie \cite{networkmag01:sanchez}. A trie is a binary tree, where a prefix is represented by a node. The value of the prefix corresponds to the path from the root of the tree to the node representing the prefix. The branching decisions are made based on the consecutive bits in the prefix. A trie is called a uni-bit trie if only one bit is used for making branching decision at a time. The prefix set in Figure \ref{fig:trie} (a) corresponds to the uni-bit trie in Figure \ref{fig:trie} (b). For example, the prefix ``010*'' corresponds to the path starting at the root and ending in node P3: first a left-turn (0), then a right-turn (1), and finally a turn to the left (0). Each trie node contains two fields: the represented prefix and the pointer to the child nodes. By using the optimization called \textit{leaf-pushing} \cite{tocs99:srinivasan}, each node needs only one field: either the pointer to the next-hop address or the pointer to the child nodes. Figure \ref{fig:trie} (c) shows the leaf-pushed uni-bit trie derived from Figure \ref{fig:trie} (b). 

Given a leaf-pushed uni-bit trie, IP lookup is performed by traversing the trie according to the bits in the IP address. When a leaf is reached, the prefix associated with the leaf is the longest matched prefix for that IP address. The time to look up a uni-bit trie is equal to the prefix length. The use of multiple bits in one scan can increase the search speed. Such a trie is called a multi-bit trie. The number of bits scanned at a time is called \textit{stride}. Some optimization schemes \cite{sigcommccr04:eatherton, toc07:kim} have been proposed to build a memory-efficient multi-bit trie. For simplicity, we consider only the leaf-pushed uni-bit trie in this paper, though our ideas are applicable to other forms of tries.

\subsubsection{Decision Tree based Packet Classification}
Multi-dimensional packet classification is one of the fundamental challenges in designing high speed routers. It enables routers to support firewall processing, Quality of Service differentiation, virtual private networks, policy routing, and other value added services. An IP packet can be classified based on a number of fields in the packet header, such as source/destination IP address, source/destination port number, type of service, type of protocol, etc. Fields are generally specified by range. When a packet arrives at a router, its header is compared against a set of rules. Each rule can have one or more fields and their associated values, a priority, and an action to be taken if matched. A packet is considered matching a rule only if it matches all the fields within that rule.

Many packet classification algorithms are based on decision trees which take the geometric view of the packet classification problem. HyperCuts \cite{sigcomm03:singh} is a representative of such algorithms. At each node of the decision tree, the search space is cut based on the information from one or more fields in the rule. HyperCuts algorithm allows cutting on multiple fields per step, resulting in a fatter and shorter decision tree.

The searching algorithm in a HyperCuts tree is simple. When a packet header arrives at the root of the tree, it will traverse the decision tree until it finds either a leaf node or a NULL node. The leaf node will represent the first matching rule, and the NULL node will indicate that no match has been found.

\subsection{Memory-Balanced Pipelines}

Pipelining can dramatically improve the throughput of tree traversal. A straightforward way to pipeline a tree is to assign each tree level to a different stage, so that a packet can be processed every clock cycle. However, as discussed earlier, this simple pipelining scheme results in unbalanced memory distribution, leading to low throughput and inefficient memory allocation. 

Basu et al. \cite{infocom03:basu} and Kim et al. \cite{toc07:kim} both reduce the memory imbalance by using variable strides to minimize the largest trie level. However, even with their schemes, the size of the memory of different stages can have a large variation. As an improvement upon \cite{toc07:kim}, Lu et al. \cite{iscc06:lu} propose a tree-packing heuristic to balance the memory further, but it does not solve the fundamental problem of how to retrieve one node's descendants which are not allocated in the following stage. Furthermore, a variable stride multi-bit trie is difficult for hardware implementation, especially if incremental updating is needed \cite{infocom03:basu}.

Baboescu et al. \cite{isca05:baboescu} propose a Ring pipeline architecture for tree-based search engines. The pipeline stages are configured in a circular, multi-point access pipeline so that the search can be initiated at any stage. A tree is split into many small subtrees of equal size. These subtrees are then mapped to different stages to create an almost balanced pipeline. Some subtrees must wrap around if their roots are mapped to the last several stages. Any incoming IP packet needs to lookup an index table to find its corresponding subtree's root which is the starting point of that search. Since the subtrees may be from different depths of the original tree, we cannot use a constant number of address bits to index the table. Thus, the index table must be built by content addressable memories (CAMs), which may result in lower speed. Though all IP packets enter the pipeline from the first stage, their lookup processes may be activated at different stages. All the packets must traverse the pipeline twice to complete the tree traversal. The throughput is thus 0.5 packets per clock cycle.

Kumar et al. \cite{ancs06:kumar} extend the circular pipeline with a new architecture called Circular, Adaptive and Monotonic Pipeline (CAMP). It uses several initial bits (i.e. initial stride) as the hashing index to partition the tree. Using the similar idea but different mapping algorithm from Ring\cite{isca05:baboescu}, CAMP creates an almost balanced pipeline as well. Unlike the Ring pipeline, CAMP has multiple entry stages and exit stages. 
To manage the access conflicts between packets from current and preceding stages, several queues are employed. Since different packets of an input stream may have different entry and exit stages, the ordering of the packet stream is lost when passing through CAMP. Assuming the packets traverse all the stages, when the packet arrival rate exceeds 0.8 packets per clock cycle, some packets may be discarded \cite{ancs06:kumar}. In other words, the worst-case throughput is 0.8 packets per clock cycle. Also in CAMP, a queue adds extra delay for each packet, which may result in out-of-order output and delay variation.


Due to the non-linear structure, neither the Ring pipeline nor CAMP in the worst case can maintain a throughput of one packet per clock cycle. Also, neither of them supports the non-blocking route update, since the ongoing update may conflict with the preceding or following packets. Our previous work \cite{hoti07:jiang} adopts an optimized linear pipeline architecture, named OLP, to achieve a high throughput of one output per clock cycle, while supporting \textit{write bubbles} \cite{infocom03:basu} for non-blocking update. By adding $nop$s (no-operations) in the pipeline, OLP offers more freedom in mapping tree nodes to pipeline stages. The tree is partitioned, and all subtrees are converted into queues and are mapped onto the pipeline from the first stage. However, in OLP, the first several stages may not be balanced, since the top levels of a tree have few nodes. 

\subsection{Discussion}
\label{sec:discussion}

State-of-the-art techniques cannot achieve perfectly balanced memory distribution, due to several constraints they place during mapping: (1) They require the tree nodes on the same level be mapped onto the same stage. (2) The mapping scheme is uni-directional: the subtrees partitioned from the original tree must be mapped in the same direction (either from the root, or from the leaves). Actually, both constraints are unnecessary. The only constraint we must obey is:

\textit{Constraint 1}: If node $A$ is an ancestor of node $B$ in a tree, then $A$ must be mapped to a stage preceding the stage to which $B$ is mapped.


This paper proposes a flexible bidirectional linear pipeline architecture which provides a unified architecture for both IP lookup and packet classification. By employing widely-used dual-port SRAMs, the architecture allows two flows from opposite directions to access the local memory in a stage at the same time. With a bidirectional fine-grained mapping scheme, a perfectly balanced memory distribution over pipeline stages is achieved. It has many desirable properties due to its linear structure: (1) the worst-case throughput of one packet per clock cycle is sustained; (2) each packet has a constant delay to go through the architecture; (3) the packet input order is maintained; (4) non-blocking update is supported, that is, while a \textit{write bubble} is inserted to update the stages, both the subsequent and the antecedent packets can perform the search as well. 


\begin{figure*}[hbt]
\centering
\subfigure[Depth-based mapping]{\includegraphics[width=3.0in]{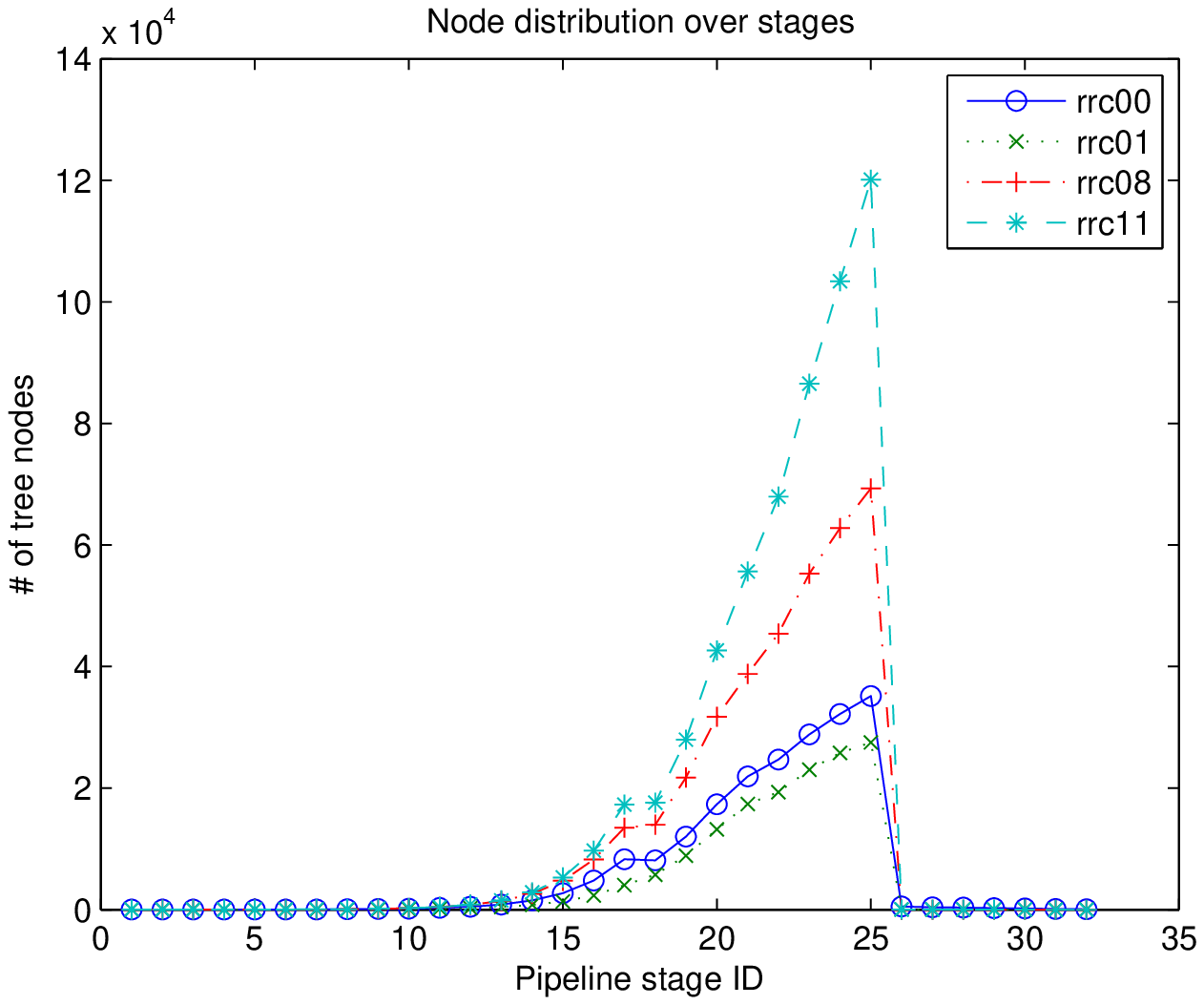}
\label{fig:ip_nod}}
\hfil
\subfigure[Height-based mapping]{\includegraphics[width=3.0in]{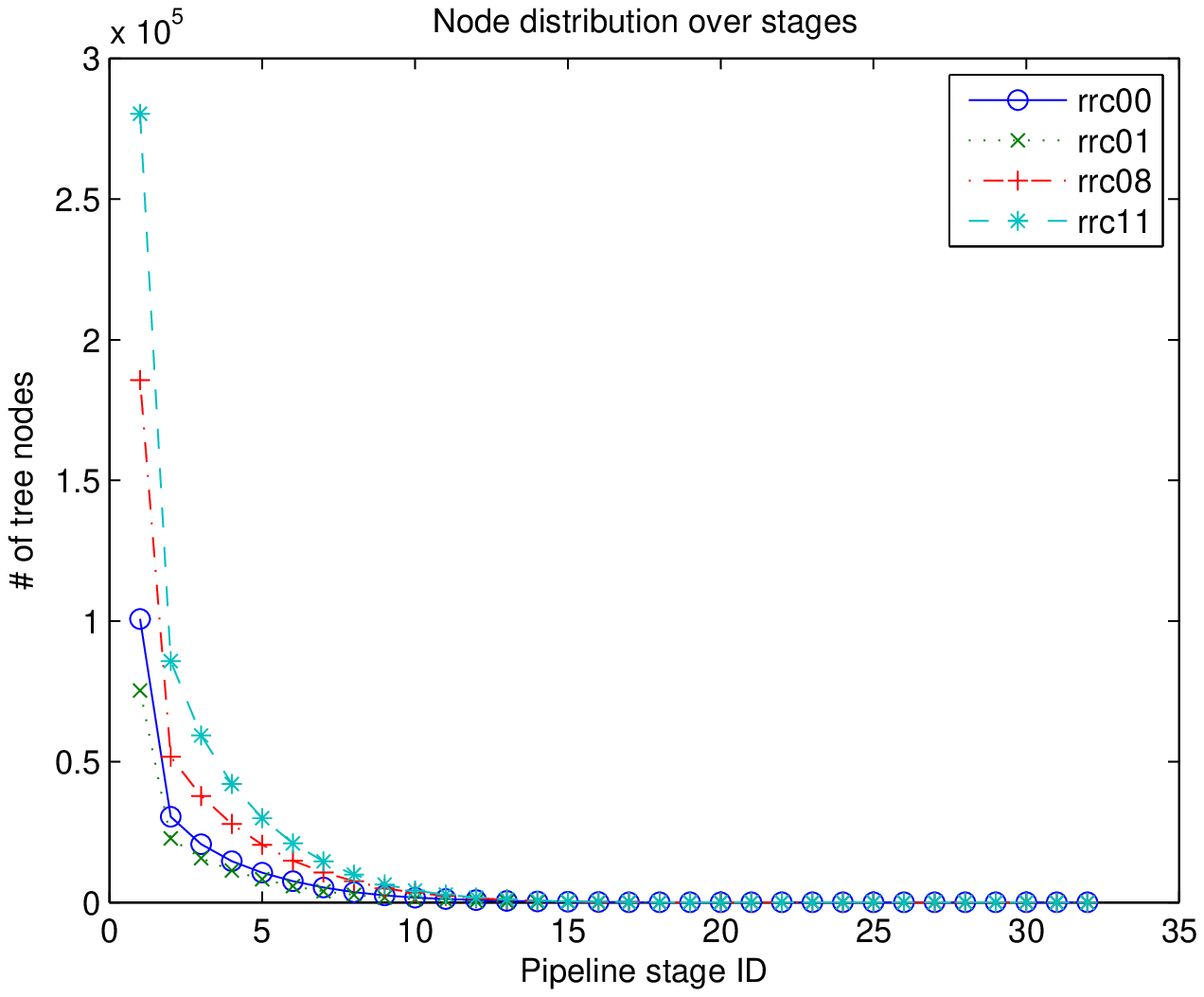}
\label{fig:ip_noh}}
\caption{Level-by-level mapping of routing tries onto 32 pipeline stages.}
\label{fig:level_mapping}
\end{figure*}

\begin{figure*}[hbt]
\centering
\subfigure[Depth-based mapping]{\includegraphics[width=3.0in]{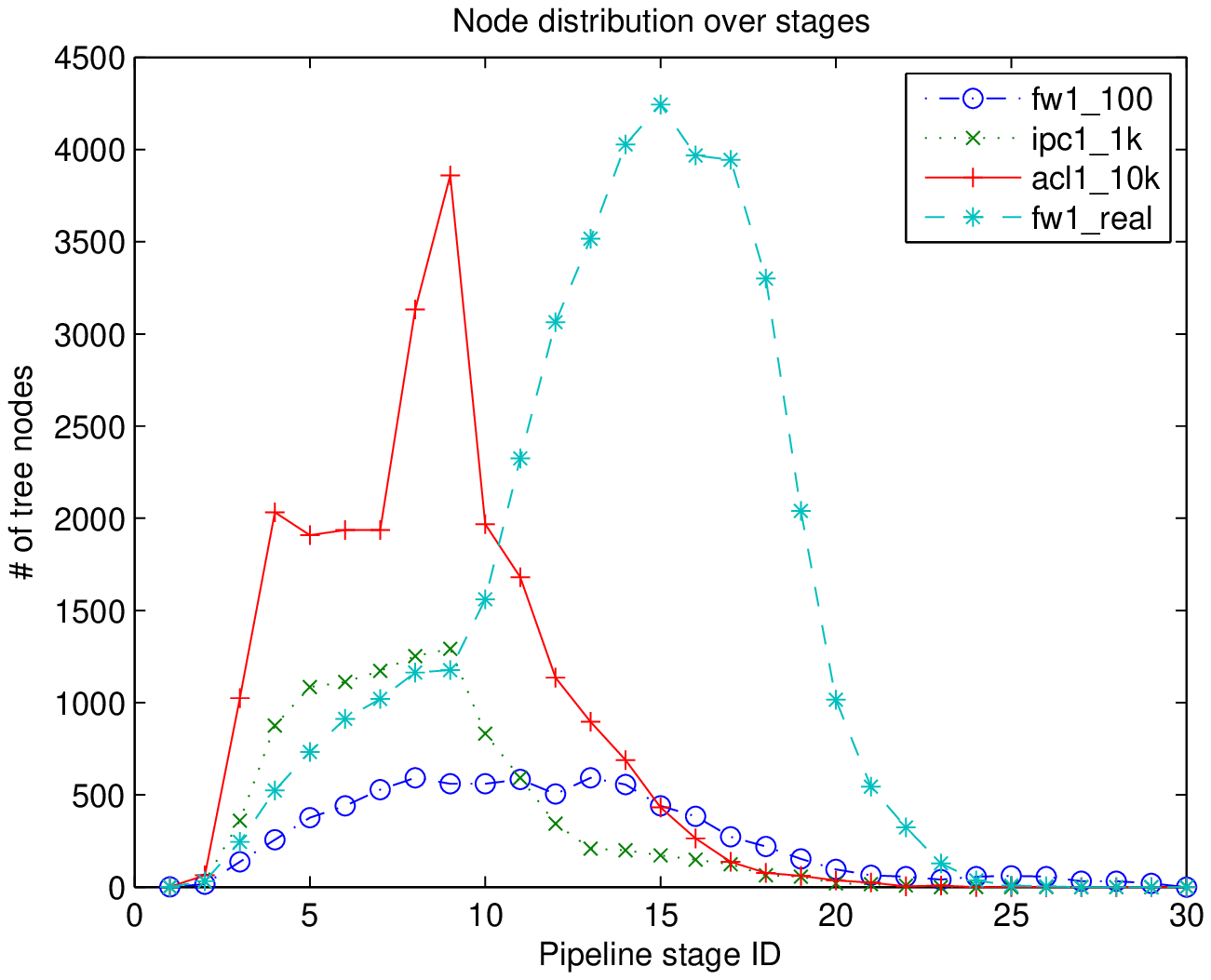}
\label{fig:pkt_nod}}
\hfil
\subfigure[Height-based mapping]{\includegraphics[width=3.0in]{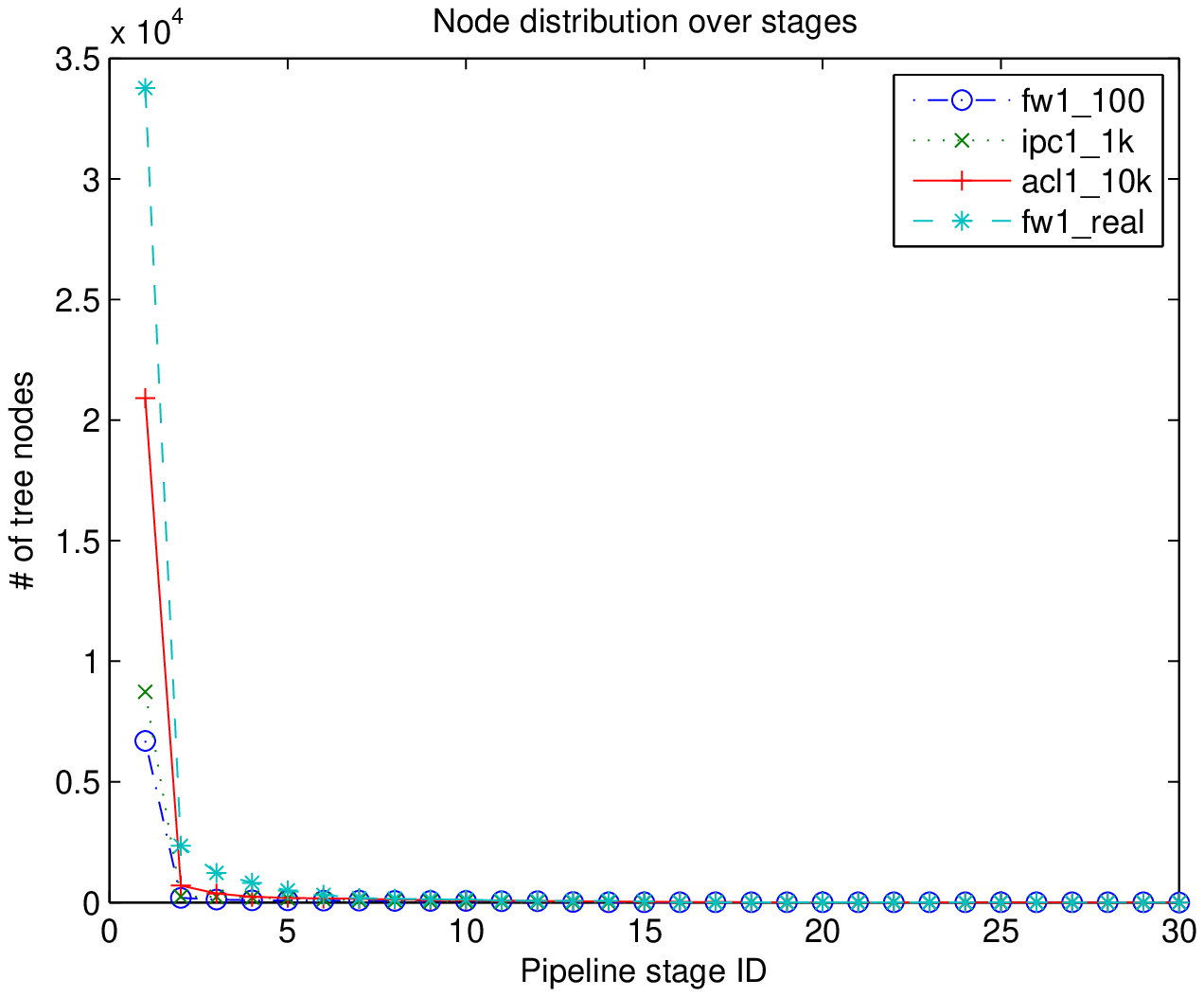}
\label{fig:pkt_noh}}
\caption{Level-by-level mapping of decision trees onto 25 pipeline stages.}
\label{fig:pkt_level_mapping}
\end{figure*}

\section{Memory Balancing}
\label{sec:membalance}

This section studies the problem of balancing the memory distribution across pipeline stages. We examine two typical mapping approaches, and then propose a novel bidirectional fine-grained mapping scheme.
First, we define the following terms.
\newtheorem{theorem}{Definition}

\begin{theorem}
The \textbf{pipeline depth} is the number of pipeline stages.
\end{theorem}
\begin{theorem}
The \textbf{depth} of a tree node is the directed distance from the tree node to the tree root. The depth of a tree refers to the maximum depth of all tree leaves.
\end{theorem}
\begin{theorem}
The \textbf{height} of a tree node is the maximum directed distance from the tree node to a leaf node. The height of a tree refers to the height of the root. In fact the depth of a tree is equal to its height.
\end{theorem}
\begin{theorem}
In depth-based (height-based) mapping, two tree nodes are said to be on the same \textbf{level} if they have the same depth (height).
\end{theorem}

\subsection{Motivation}
\label{sec:motivation}

The most straightforward mapping scheme is depth-based mapping \cite{infocom03:basu}, where the tree nodes with the same depth are mapped onto the same stage. In this scheme, the first stage always has one tree node i.e. the tree root. All the packets enter the pipeline from the first stage.
Another level-by-level mapping scheme is height-based mapping \cite{sigcomm05:hasan}, where the tree nodes with the same height are mapped onto the same stage. In this scheme, all tree leaves are mapped onto the first stage, and the tree root is mapped onto the last stage. All the packets enter the pipeline from the last stage. 

We study the effectiveness of the above two mapping schemes by conducting experiments on four representative routing tables $rrc00$, $rrc01$, $rrc08$ and $rrc11$ collected from \cite{ripe:ris}. We also collected four rule sets, $fw1\_100$, $ipc1\_1k$, $acl1\_10k$ and $fw1\_real$, from \cite{ruleset} and built them into decision trees using the HyperCuts algorithm \cite{sigcomm03:singh}. According to Figures \ref{fig:level_mapping}(a-b) and Figures \ref{fig:pkt_level_mapping}(a-b), for both trie-based IP lookup and decision tree based packet classification, the node distribution across the stages is extremely unbalanced after using either the depth-based or the height-based mapping. 

\begin{figure*}[htb]
\centering
\includegraphics[width=6.0in]{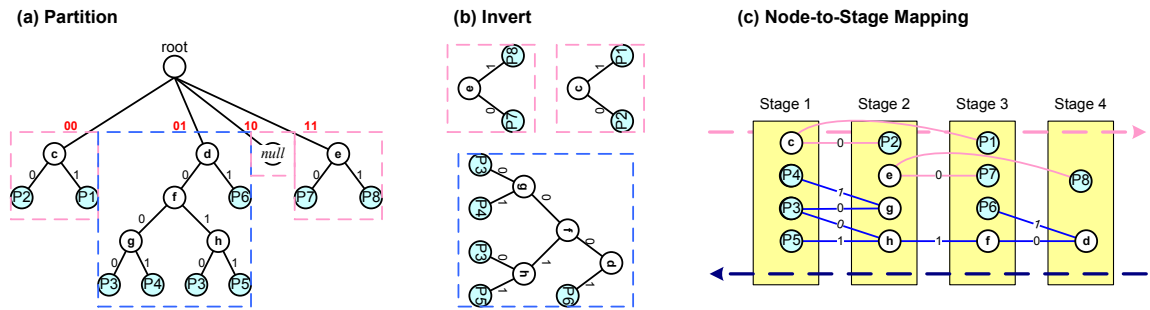}
\caption{Bidirectional fine-grained mapping for the trie in Figure \ref{fig:trie}.}
\label{fig:biolp_scheme}
\end{figure*}

We say the depth-based mapping is \textit{forward} mapping since the mapping is begun from the root, while the height-based mapping is \textit{reverse} mapping since it is begun from the leaves. Intuitively it will be more balanced if the two mapping schemes can be combined in an effective way.


%

%

\subsection{Bidirectional Fine-grained Mapping}

To achieve a perfectly balanced memory distribution over the stages, we propose a bidirectional fine-grained mapping scheme\footnote{For simplicity, in this section we describe our scheme for the trie only. The scheme can be easily extended for the decision tree.}, as shown in Figure \ref{fig:biolp_scheme}. The main ideas are (1) \textbf{fine-grained mapping}: allowing two trie nodes on the same trie level to be mapped onto different stages; and (2) \textbf{bidirectional mapping}: allowing two subtrees to be mapped onto different directions. However, there are several issues to be addressed:

\begin{itemize}
	\item Partition the entire tree so that we can have several subtrees to be mapped in different directions.
	\item Decide which subtree(s) should be inverted and mapped on the reverse direction.
	\item Adapt and combine the depth-based and the height-based mapping schemes at each step.
\end{itemize}

\subsubsection{Tree Partitioning}
\label{sec:partition}

We use prefix expansion \cite{tocs99:srinivasan} to partition the tree. Several initial bits are used as the index to partition the tree into many disjoint subtrees. According to \cite{sigcomm01:baboescu} and our observation on the collected routing tables, few prefixes in real-life routing tables are shorter than 16. Hence, there will be little prefix duplication when we use fewer than 16 initial bits to expand the prefixes.

\subsubsection{Subtree Inversion}
\label{sec:invert}

In a trie, there are few nodes at the top levels while there are a lot of nodes at the leaf level. Hence, we can invert some subtrees so that their leaf nodes are mapped onto the first several stages. We propose several heuristics to select the subtrees to be inverted: 
\begin{compactenum}
	\item \textit{Largest leaf}: The subtree with the most number of leaves is preferred. This is straightforward since we need enough nodes to be mapped onto the first several stages.
	\item \textit{Least height}: The subtree of shortest height is preferred. Due to \textit{Constraint 1}, a subtree with a larger height has less flexibility to be mapped onto pipeline stages.
	\item \textit{Largest leaf per height}: This is a combination of the previous two heuristics, by dividing the number of leaves of a subtree by its height.
	\item \textit{Least average depth per leaf}: Average depth per leaf is the ratio of the sum of the depth of all the leaves to the number of leaves. This heuristic prefers a more balanced subtree. 
\end{compactenum}

Algorithm \ref{alg:invert} finds the subtrees to be inverted, where $IFR$ denotes the \textit{inversion factor}. A larger inversion factor results in more subtrees to be inverted. When the inversion factor is 0, no subtree is inverted. When the inversion factor is close to the pipeline depth, all subtrees are inverted. The complexity of this algorithm is $O(K)$ where $K$ denotes the total number of subtrees.

\begin{algorithm}[htb]
\caption{Selecting the subtree to be inverted}
\label{alg:invert}
\begin{algorithmic}[1]
\REQUIRE $K$ subtrees.
\ENSURE $V$ subtrees to be inverted.
\STATE $N =$ total $\#$ of tree nodes of all subtrees, $H = \#$ of pipeline stages, $V = 0$, $W = K$.
\WHILE{$V < K$ AND $W < IFR \times \left\lceil N / H\right\rceil$}
\STATE Based on the chosen heuristic, select one subtree from those not inverted.
\STATE $V = V + 1$, $W = W - 1 + \#$ of leaves of the selected subtree.
\ENDWHILE
\end{algorithmic}
\end{algorithm}

\subsubsection{Mapping Algorithm}

Now we have two sets of subtrees. Those subtrees which are mapped from roots are called the \textit{forward subtrees}, while the others are called the \textit{reverse subtrees}. We use a bidirectional fine-grained mapping algorithm (Algorithm \ref{alg:biolp}). The nodes are popped out of the $ReadyList$ in the decreasing order of their priority. The priority of a trie node is defined as its height if the node is in a forward subtree, and its depth if in a reverse subtree. The node whose priority is equal to the number of the remaining stages is regarded as a \textit{critical} node. If such a node is not mapped onto the current stage, some of its descendants (if in a forward subtree) or ascendants (if in a reverse subtree) can not be mapped later. For the forward subtrees, a node is pushed into the $NextReadyList$ immediately after its parent is popped. For the reverse subtrees, a node will not be pushed into the $NextReadyList$ until all its children are popped. The complexity of this mapping algorithm is $O(HN)$ where $H$ denotes the pipeline depth and $N$ the total number of trie nodes.

\begin{algorithm}[htb]
\caption{Bidirectional fine-grained mapping}
\label{alg:biolp}
\begin{algorithmic}[1]
\REQUIRE $K$ forward subtrees and $V$ reverse subtrees.
\ENSURE $H$ stages with mapped nodes.
\STATE Create and initialize two lists: $ReadyList = \phi$ and $NextReadyList = \phi$.
\STATE $R_n = \#$ of remaining nodes, $R_h = \#$ of remaining stages = $H$.
\STATE Push the roots of the forward subtrees and the leaves of the reverse subtrees into $ReadyList$.
\FOR{$i = 1$ to $H$}
\STATE $M_i = 0$, $Critical = FALSE$.
\STATE Sort the nodes in $ReadyList$ in the decreasing order of the node priority.
\WHILE{$Critical = TRUE$ or ($M_i < \left\lceil R_n / R_h\right\rceil$ and $Readylist \neq \phi$)}
\STATE Pop node from $ReadyList$ and map into Stage $i$. 
\IF{The node is in forward subtrees}
\STATE The popped node's children are pushed into $NextReadyList$.
\ELSIF{All children of the popped node's parent have been mapped}
\STATE The popped node's parent is pushed into $NextReadyList$.
\ENDIF
\STATE $Critical = FALSE$.
\IF{There exists a node $N_c \in ReadyList$ and the priority of $N_c$ $>= R_h - 1$}
\STATE $Critical = TRUE$.
\ENDIF
\ENDWHILE
\STATE $R_n = R_n - M_i$, $R_h = R_h - 1$.
\STATE Merge the $NextReadyList$ to the $ReadyList$.
\ENDFOR
\end{algorithmic}
\end{algorithm}

\begin{figure*}[bht]
\centering
\includegraphics[width=5.0in]{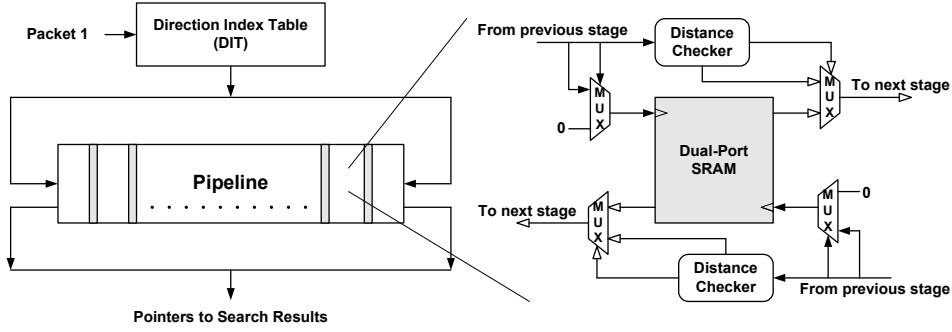}
\caption{Block diagram of the basic architecture.}
\label{fig:basic_arch}
\end{figure*}

The effectiveness of the bidirectional mapping scheme is evaluated in Section \ref{sec:performance}.


\section{Hardware Architecture}
\label{sec:arch}

To enable the bidirectional fine-grained mapping scheme, we develop a bidirectional linear pipeline architecture based on dual-port SRAMs\footnote{Dual-port SRAMs have been standard components in many devices such as FPGAs \cite{xilinx:virtex}.}, as shown in Figure \ref{fig:basic_arch}.

\subsection{Overview}

There is one Direction Index Table (DIT), which stores the relationship between the subtrees and their mapping directions: forward or reverse. For any arriving packet $p$, the initial bits of its IP address are used to lookup the DIT and retrieve the information about its corresponding subtree $ST(p)$. The information includes (1) the distance to the stage where the root of $ST(p)$ is stored, (2) the memory address of the root of $ST(p)$ in that stage, and (3) the mapping direction of $ST(p)$ which leads the packet to different entrance of the pipeline. For example, in Figure \ref{fig:basic_arch}, if the mapping direction is forward, the packet is sent to the leftmost stage of the pipeline. Otherwise, the packet is sent to the rightmost stage. 

Once its direction is known, the packet will go through the entire pipeline in that direction. The pipeline is configured as a dual-entrance bidirectional linear pipeline. At each stage, the memory has dual Read/Write ports so that the packets from both directions can access the memory simultaneously. The content of each entry in the memory includes (1) the memory address of the child node and (2) the distance to the stage where the child node is stored. If the distance value is zero, the memory address of its child node will be used to index the memory in the next stage to retrieve the child node content. Otherwise, the packet will pass that stage without any operation but decrement its distance value by one.

\subsection{Incremental Route Updates}

We update the memory in the pipeline by inserting \textit{write bubbles} \cite{infocom03:basu}. The new content of the memory is computed offline. When an update is initiated, a write bubble is inserted into the pipeline. The direction of write bubble insertion is determined by the direction of the subtree that the write bubble is going to update. Each write bubble is assigned an ID. There is one write bubble table in each stage. It stores the update information associated with the write bubble ID. When it arrives at the stage prior to the stage to be updated, the write bubble uses its ID to lookup the write bubble table. Then it retrieves (1) the memory address to be updated in the next stage, (2) the new content for that memory location, and (3) a write enable bit. If the write enable bit is set, the write bubble will use the new content to update the memory location in the next stage. 


Since the subtrees mapped onto the two directions are disjoint, a write bubble inserted from one direction will not contaminate the memory content for the search from the other direction. Also, since the pipeline is linear, all packets preceding or following the write bubble can perform their searches while the write bubble performs an update.

\subsection{Throughput Improvement by Caching}

In the above architecture shown in Figure \ref{fig:basic_arch}, at most two packets are allowed to enter the pipeline at the same time. The throughput can be 2 packets per clock cycle (PPC) only if the two packets are in the two distinct directions. Usually, such a traffic balancing cannot be guaranteed in reality. Thus, the throughput is lower than 2 PPC when we insert 2 packets in one clock cycle.

\begin{figure}[thb]
\centering
\includegraphics[width=3.5in]{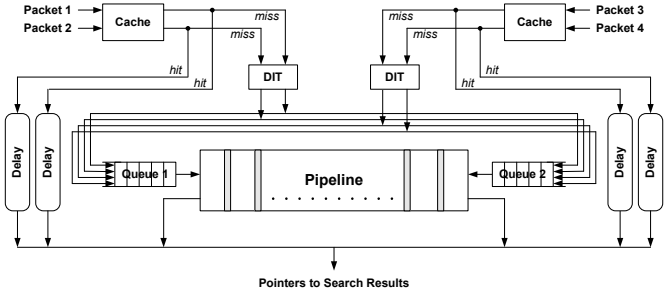}
\caption{Block diagram of the enhanced architecture.}
\label{fig:enhance_arch}
\end{figure}

On the other hand, Internet traffic contains a great amount of locality due to the TCP mechanism and application characteristics \cite{infocom08:jiang}. As shown in Figure \ref{fig:enhance_arch}, some small caches can be added into the architecture to exploit the Internet traffic locality. The most recently searched packets are cached. Any arriving packet accesses the cache first. If a cache hit happens, the packet will skip traversing the pipeline. Otherwise, it needs to go through the pipeline. The cache can be organized in any associativity. We use full associativity as the default. For IP lookup, only the destination IP of the packet is used to index the cache, while for packet classification, multiple fields of the packet header may be used. The cache update will be triggered, either when there is a route update that is related to some cached entry, or after a packet that previously had a cache miss retrieves its search result from the pipeline. Any replacement algorithm can be used to update the cache. The Least Recently Used (LRU) algorithm is used as the default.

%

%

\section{Performance Evaluation}
\label{sec:performance}

This section evaluates the effectiveness of the proposed scheme and the performance of the proposed architecture. At first, we examine the memory balancing by using the bidirectional fine-grained mapping scheme. Then, we measure the throughput using real-life traffic traces. All experiments are based on simulation.

\begin{figure*}[htb]
\centering
\subfigure[Largest leaf]{\includegraphics[width=3.0in]{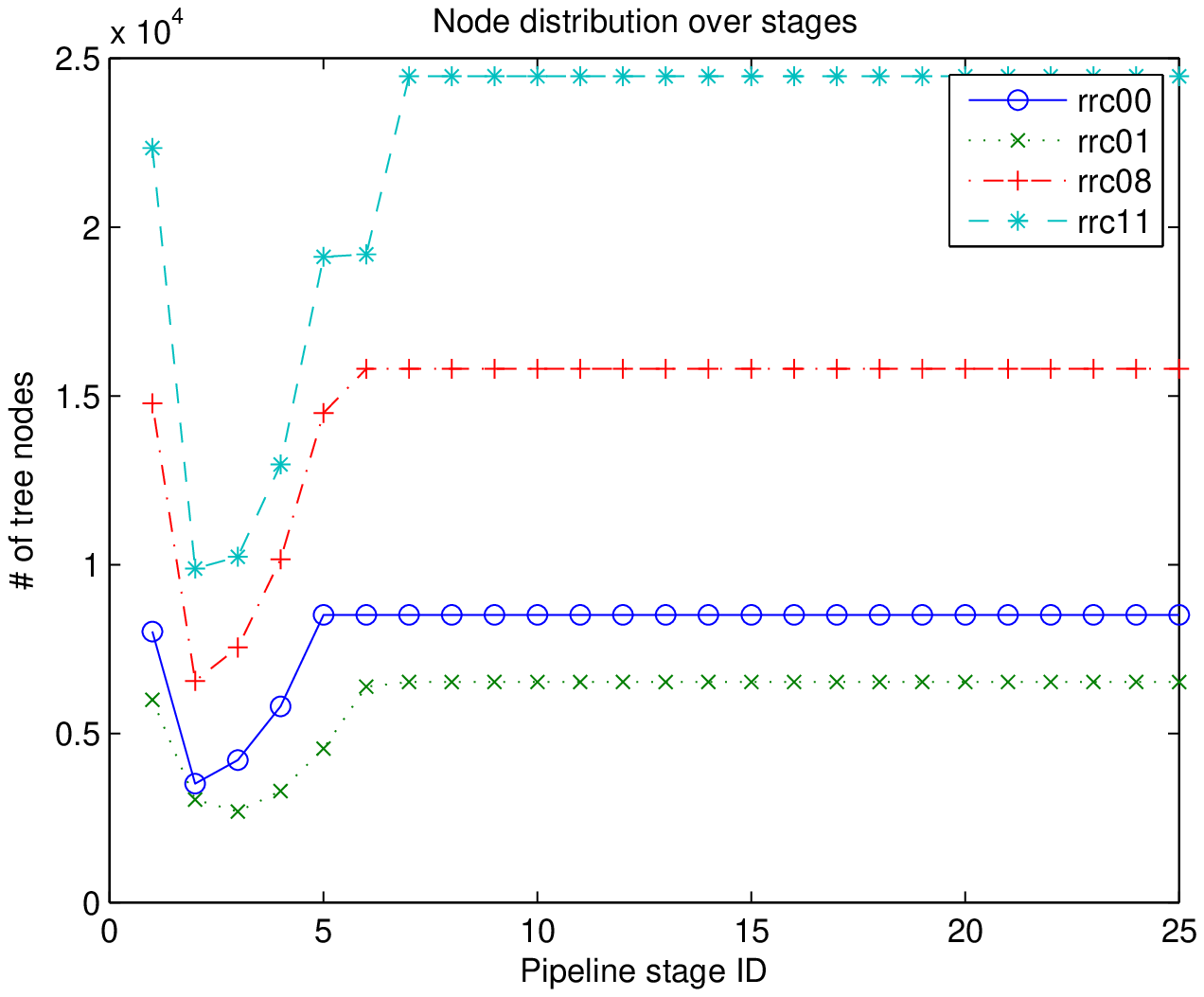}
\label{fig:ip_nos_rf1_largeleaf}}
\hfil
\subfigure[Least height]{\includegraphics[width=3.0in]{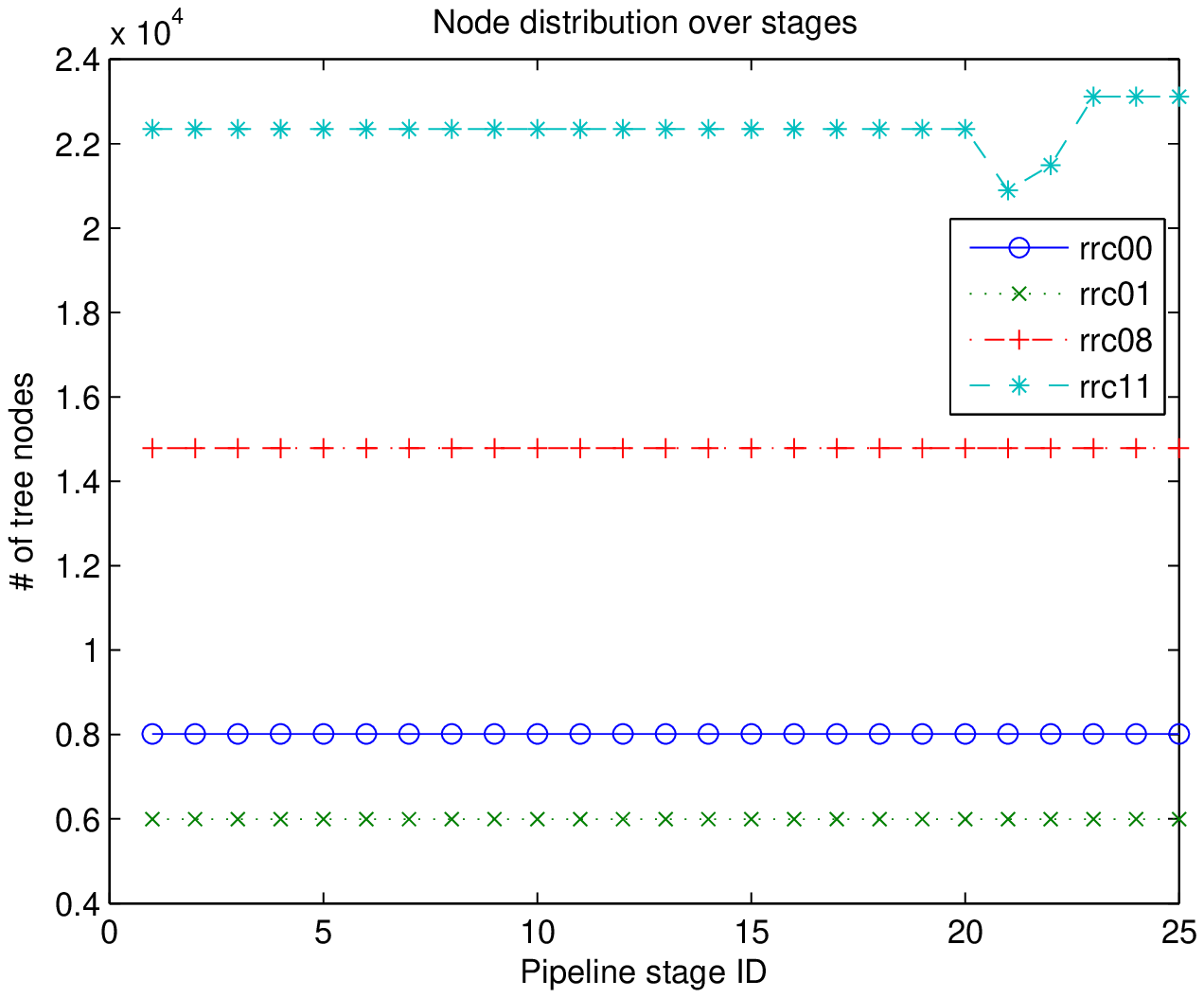}
\label{fig:ip_nos_rf1_leastheight}}
\vfil
\subfigure[Largest leaf per height]{\includegraphics[width=3.0in]{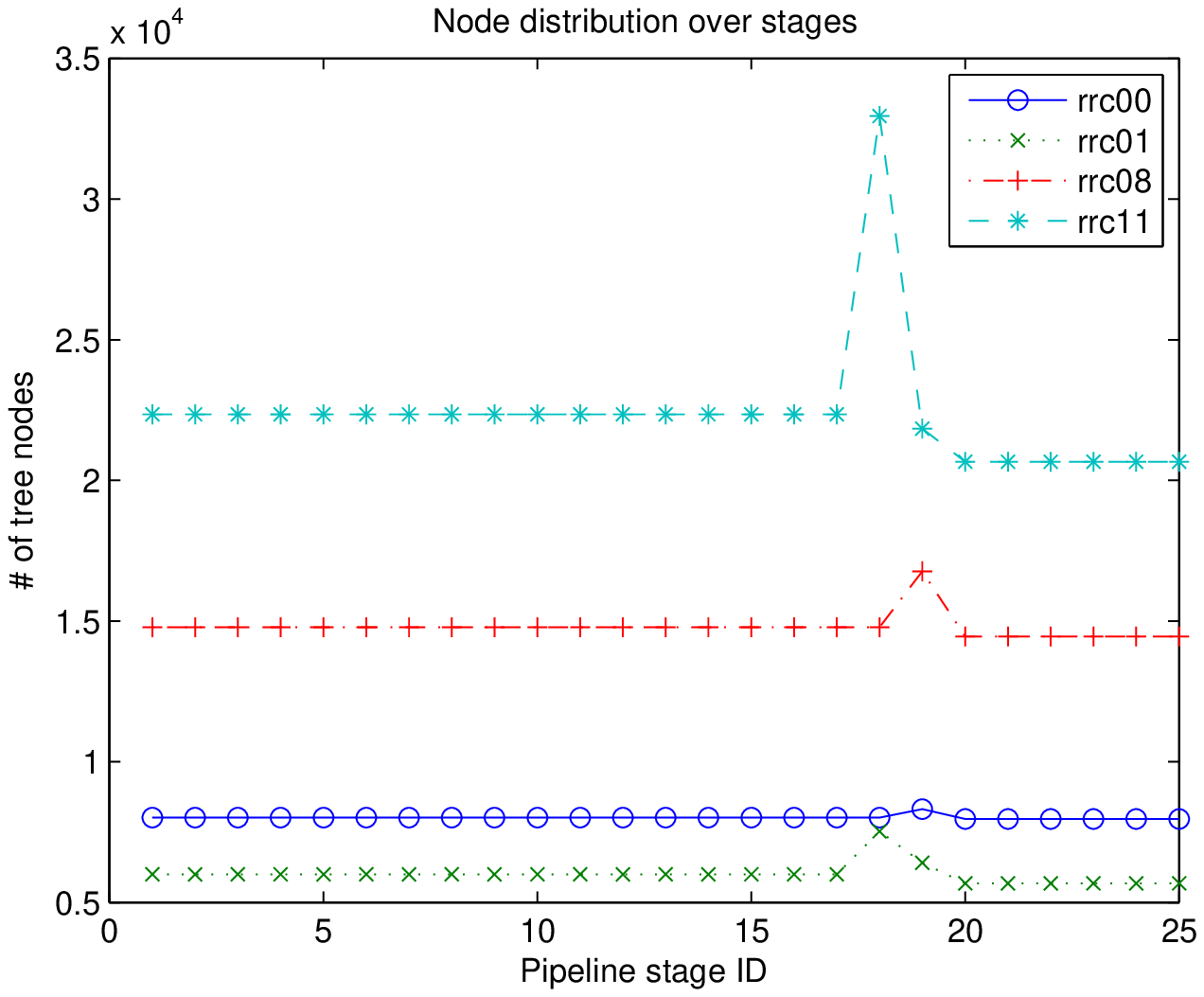}
\label{fig:ip_nos_rf1_largelph}}
\hfil
\subfigure[Least average depth per leaf]{\includegraphics[width=3.0in]{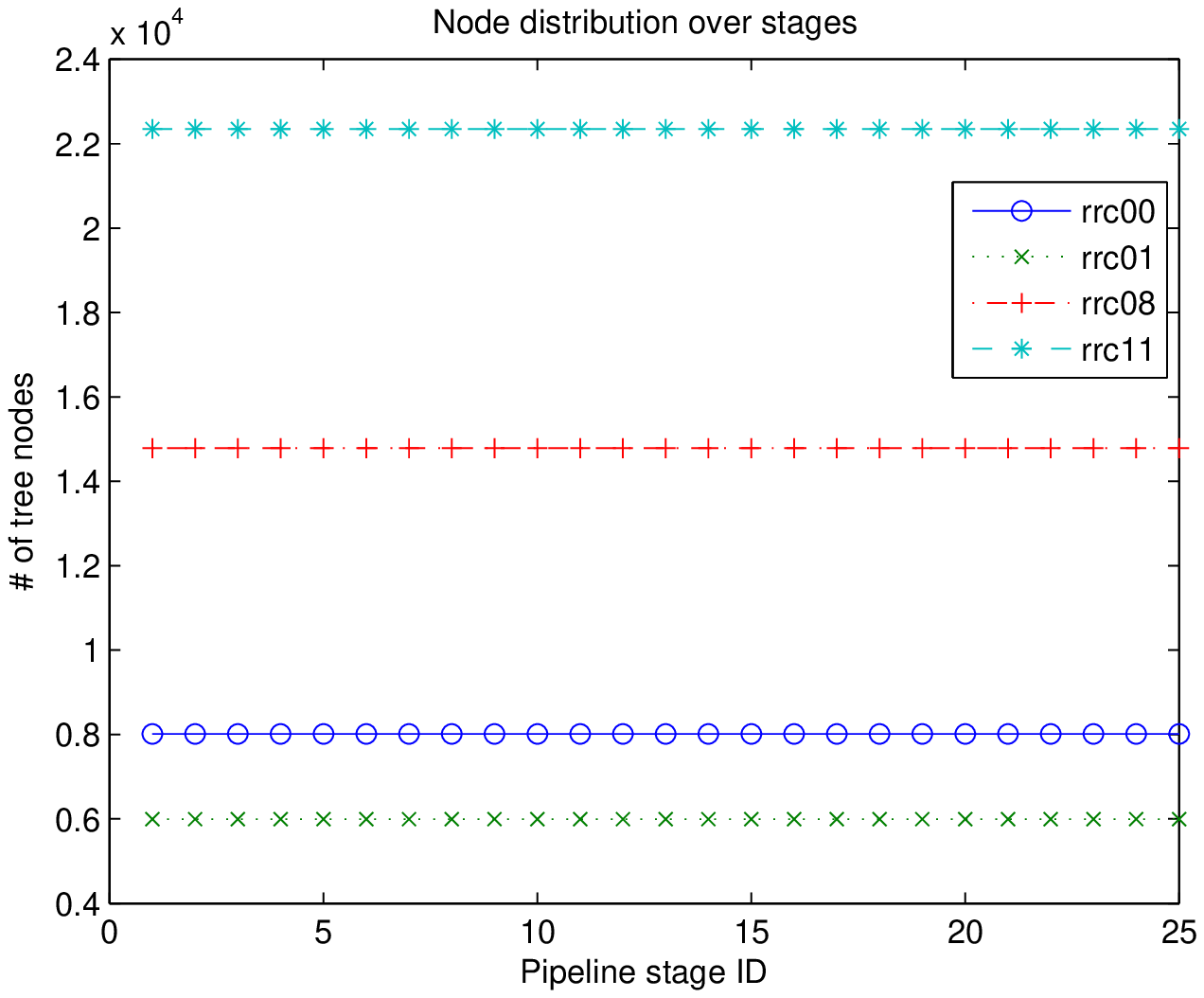}
\label{fig:ip_nos_rf1_dpl}}
\caption{Bidirectional fine-grained mapping with different heuristics. (Inversion factor = 1)}
\label{fig:biolp_heuristic}
\end{figure*}

\begin{figure*}[htb]
\centering
\subfigure[Inversion factor = 0]{\includegraphics[width=3.0in]{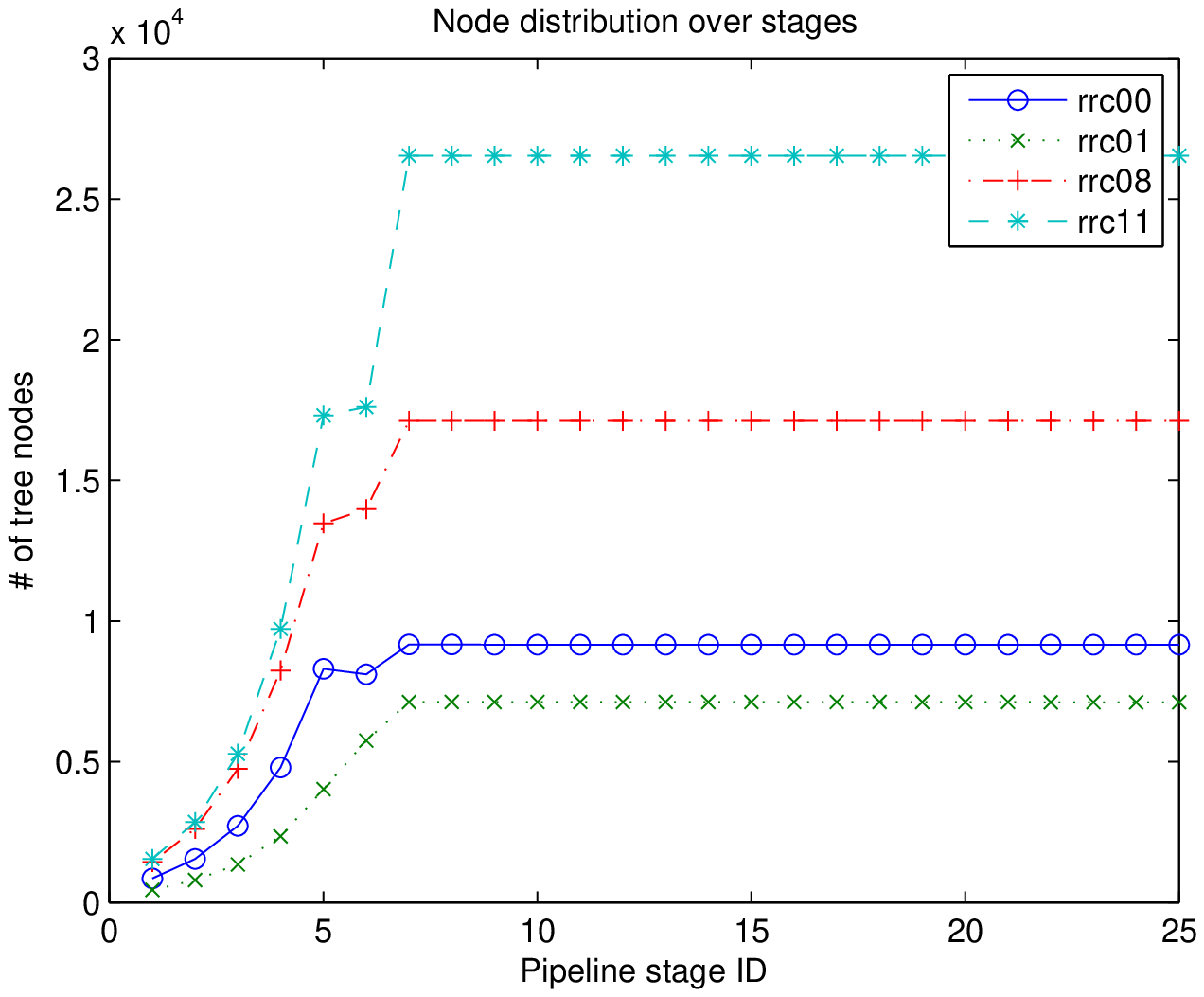}
\label{fig:ip_topdown}}
\hfil
\subfigure[Inversion factor = 4]{\includegraphics[width=3.0in]{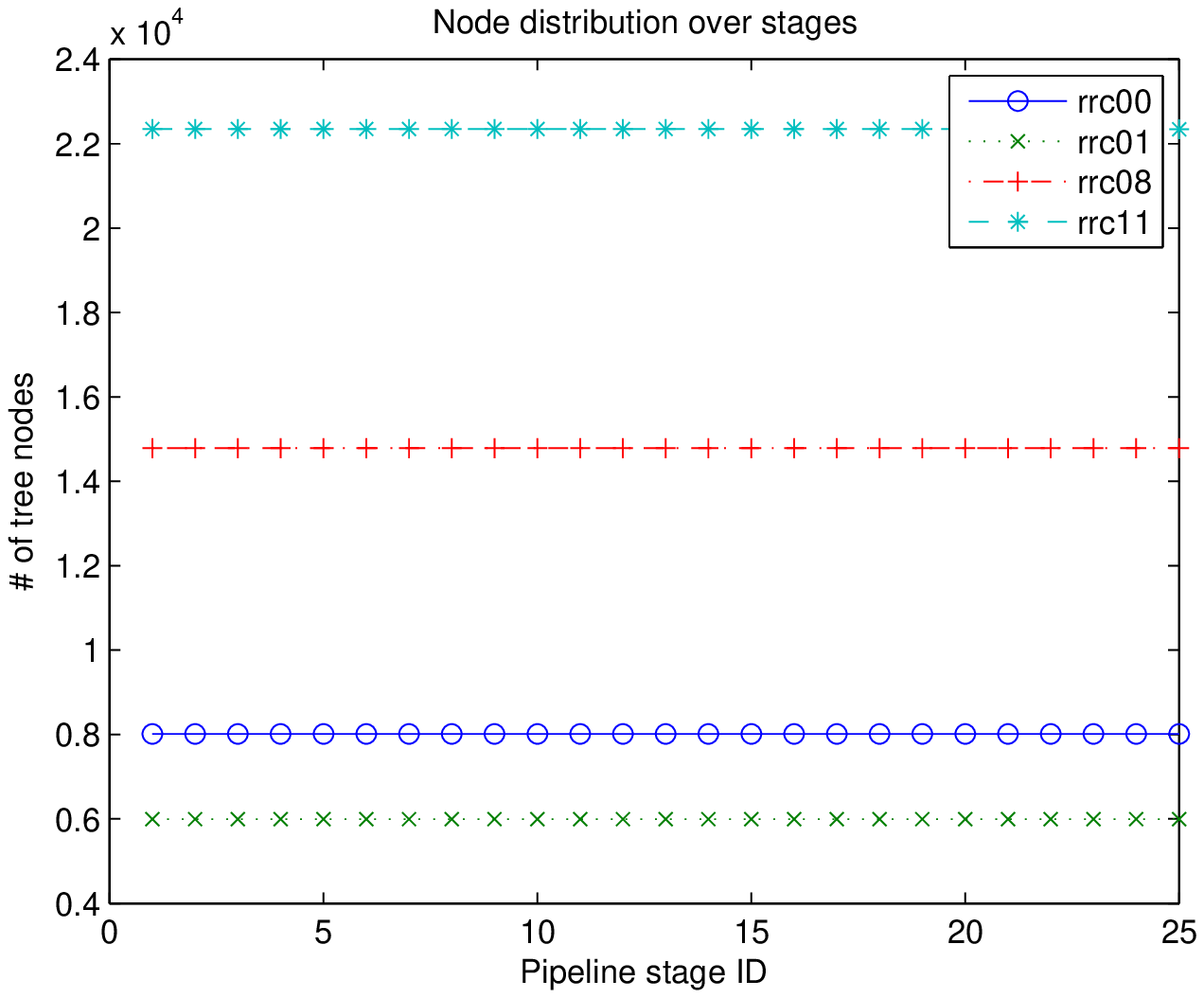}
\label{fig:ip_nos_rf4_largeleaf}}
\vfil
\subfigure[Inversion factor = 8]{\includegraphics[width=3.0in]{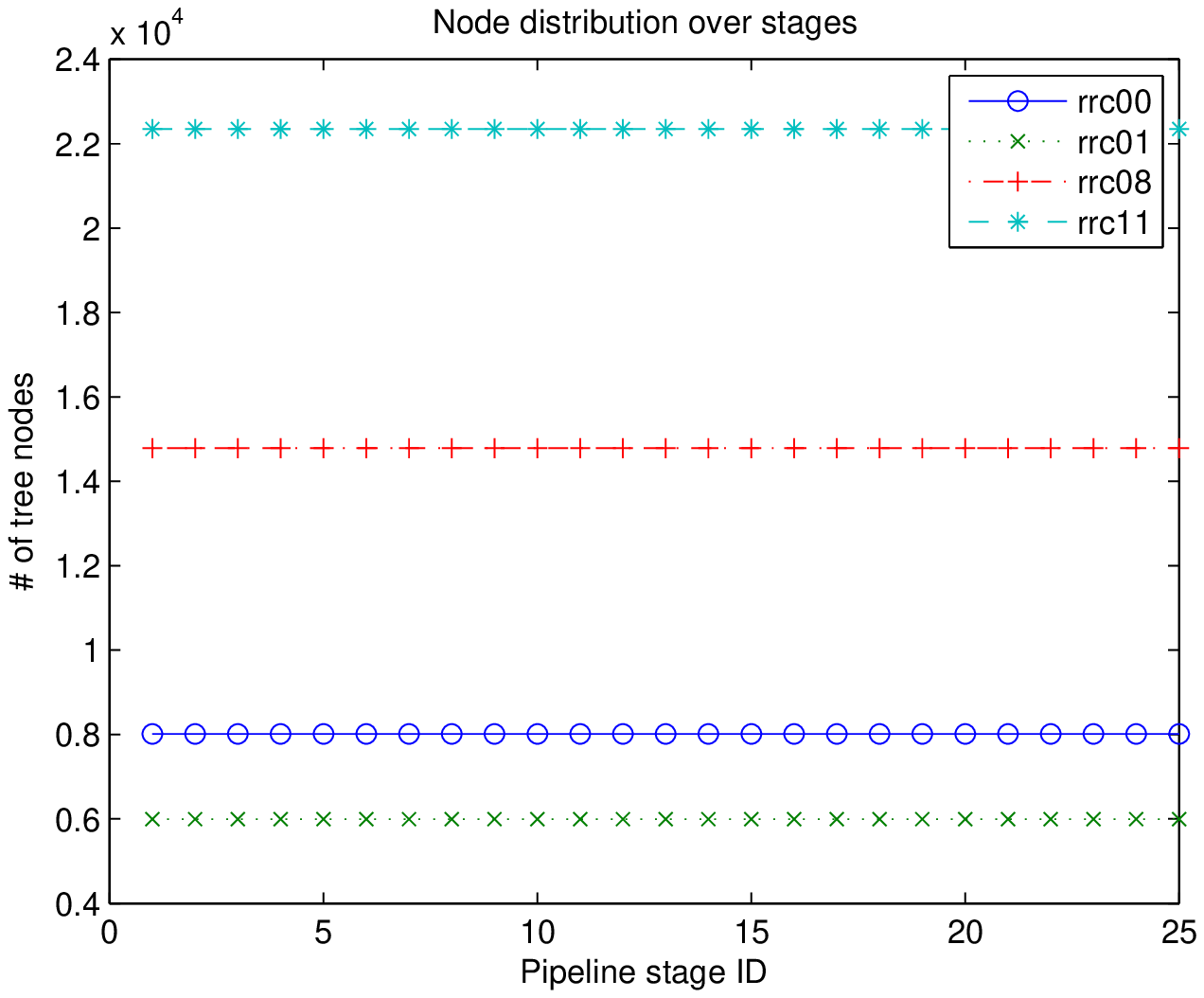}
\label{fig:ip_nos_rf8_largeleaf}}
\hfil
\subfigure[Inversion factor = 12]{\includegraphics[width=3.0in]{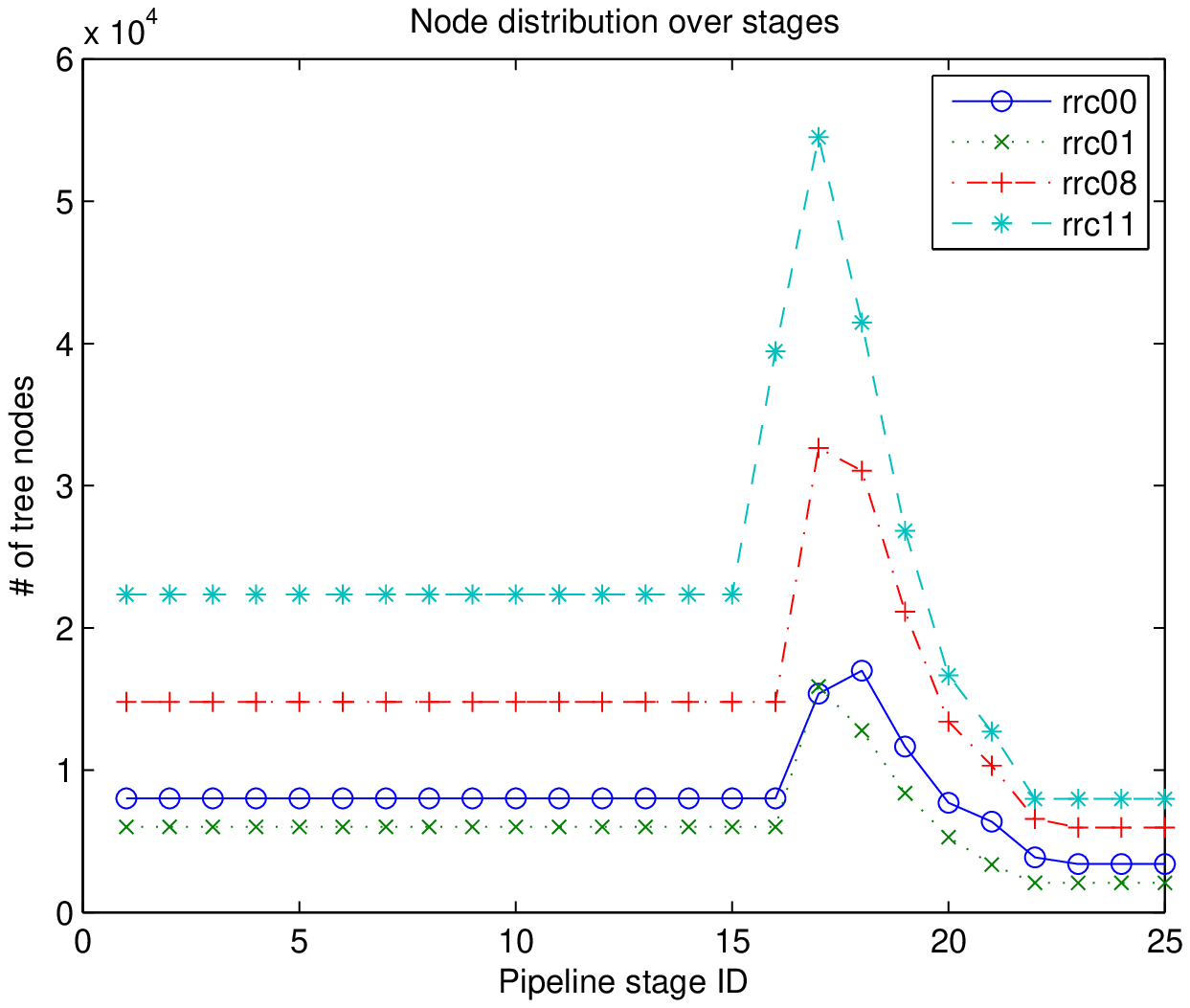}
\label{fig:ip_nos_rf12_largeleaf}}
\caption{Bidirectional fine-grained mapping with various inversion factors. (\textit{Largest leaf} heuristic)}
\label{fig:biolp_ifr}
\end{figure*}

\subsection{Memory Balancing}

At first, we conducted the experiments on the four routing tables given in Section \ref{sec:motivation} to examine the effectiveness of the bidirectional fine-grained mapping scheme. We used various inversion heuristics and inversion factor to evaluate their impacts. In these experiments, the number of initial bits used for partitioning the trie is 12. Then, with appropriate parameter setting, we conducted the experiments on the four 5-tuple rule sets to verify the effectiveness of our scheme for decision trees.  

\subsubsection{Impact of the inversion heuristics} 

As discussed in Section \ref{sec:invert}, we have four different heuristics to invert subtrees. Now we examine their performance and obtain the results shown in Figure \ref{fig:biolp_heuristic}. The value of the inversion factor is set to 1. According to the results, the \textit{least average depth per leaf} heuristic has the best performance. It shows that, when we have a choice, a balanced subtree should be inverted. This can be explained that a balanced subtree has many nodes not only at the leaf level but also at the lower levels, which can help balance not only the first stage but also the first several stages.

\subsubsection{Impact of the inversion factor}

Using the \textit{largest leaf} heuristic, we changed the value of the inversion factor. The results are shown in Figure \ref{fig:biolp_ifr}. 
When the inversion factor is 0, the bidirectional mapping becomes fine-grained forward mapping only. The mapping turns to be fine-grained reverse mapping when the inversion factor is close to the pipeline depth so that all subtrees are inverted. 


\subsubsection{Short Summary}

According to the above results for trie-based IP lookup, the bidirectional fine-grained mapping scheme can achieve a perfectly balanced memory distribution over the pipeline stages, by either using an appropriate inversion heuristic or adopting an appropriate inversion factor. This also shows that the architecture is flexible that it offers a large design space for adapting to different routing tables with various prefix distribution. In fact we conducted more experiments on 16 routing tables collected from \cite{ripe:ris} and obtained similar results as are presented.

\subsubsection{Applying onto Decision Trees}

\begin{figure}[htb]
\centering
\includegraphics[width=3.0in]{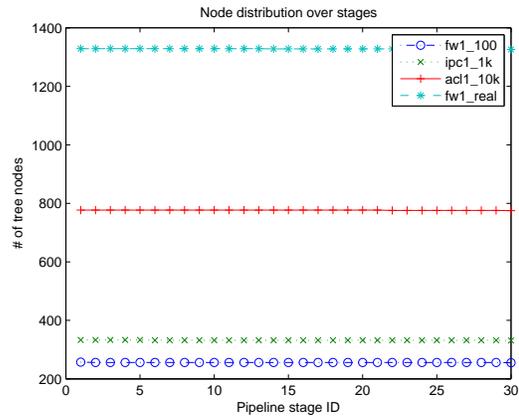}
\caption{Bidirectional fine-grained mapping for decision trees. (\textit{Largest leaf} heuristic; Inversion factor = 1)}
\label{fig:pkt_nos_rf1_largeleaf}
\end{figure}

Comparing Figures \ref{fig:level_mapping}(a-b) with Figures \ref{fig:pkt_level_mapping}(a-b), we can find that the characteristics of the decision trees are distinct from those of the routing tries. The node distribution of the decision trees after the depth-based mapping is somewhat different from that of routing tries. There are quite a lot of nodes in the first several stages, so that we can invert few subtrees in the bidirectional mapping. Also, in HyperCuts, each step from a node to its children is a multi-dimensional cut, rather than a bit scan. Hence, we cannot use prefix expansion to partition the tree. Instead, we use only the first cut to partition the tree. Figure \ref{fig:pkt_nos_rf1_largeleaf} shows the results for the bidirectional mapping of the decision trees. Compared to Figure \ref{fig:ip_nos_rf1_largeleaf}, which uses the same setting but does not achieve a balanced distribution, Figure \ref{fig:pkt_nos_rf1_largeleaf} exhibits a perfectly balanced node distribution over stages.

\subsection{Throughput Improvement}

We used real-life Internet traffic traces to evaluate the throughput performance of the proposed architecture. Two anonymized real-life traces were collected from \cite{trace:nlanr}. Their information is listed in Table \ref{tb:traces}. Due to the unavailability of public IP traces associated with their corresponding routing tables, we generated the routing tables by extracting the unique destination IP addresses from the traces. 

\begin{table}[bht]
\caption{IP header traces}
\label{tb:traces}
\vspace{0.1in}
\begin{center}
\begin{tabular}{|c|c|c|}
\hline
Trace & \# of packets & \# of IPs \\
\hline
\hline
APTH: AMP-1110523221-1 & 769100 & 17628 \\
\hline
IPLS: I2A-1091235138-1 & 1821364 & 15791\\
\hline
\end{tabular}
\end{center}
\end{table}

The major parameters of the architecture include the input width, i.e the number of parallel inputs, denoted $P$; the pipeline depth, denoted $H$; the queue size, i.e. the maximum number of packets allowed to be stored in a queue, denoted $Q$; and the cache size, i.e. the maximum number of packets allowed to be cached, denoted $C$. In these experiments, the default setting for the architecture parameters was $P = 4, H = 25, Q = 2, C = 160$.

The performance metric is the throughput in terms of the number of packets processed per clock cycle (PPC). Note that in $P$-width architecture, the throughput $\leq P$.

\subsubsection{Impact of the input width}

\begin{figure}[htb]
\centering
\includegraphics[width=3.0in]{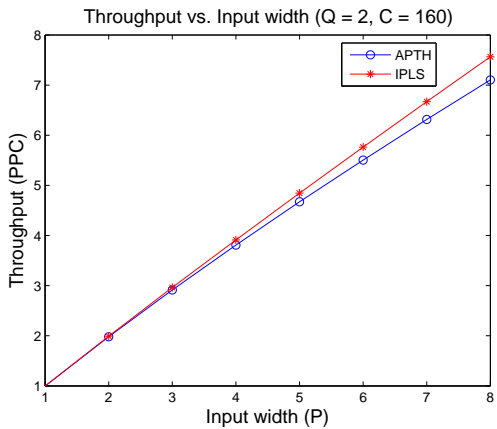}
\caption{Throughput vs. Input width. ($H = 25, Q = 2, C = 160$.)}
\label{fig:ppc_iw}
\end{figure}

We increased the input width, and observed the throughput scalability. Figure \ref{fig:ppc_iw} shows, with caching, the throughput scaled well with the input width, especially when $P \leq 4$.

\subsubsection{Impact of the cache size and the queue size}

\begin{figure}[htb]
\centering
\includegraphics[width=3.0in]{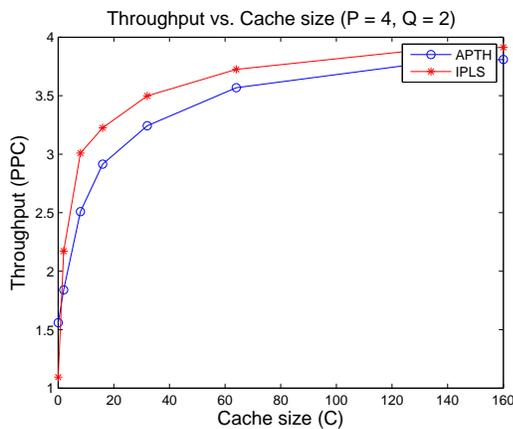}
\caption{Throughput vs. Cache size. ($P = 4, H = 25, Q = 2$.)}
\label{fig:ppc_cs}
\end{figure}

\begin{figure}[htb]
\centering
\includegraphics[width=3.0in]{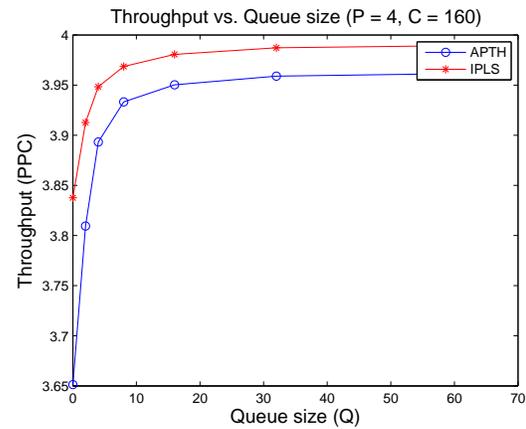}
\caption{Throughput vs. Queue size. ($P = 4, H = 25, C = 160$.)}
\label{fig:ppc_qs}
\end{figure}

We evaluated the impact of the cache size and the queue size, respectively, on the throughput. The results are shown in Figures \ref{fig:ppc_cs} and \ref{fig:ppc_qs}. Caching is efficient in improving the throughput. With only 1\% of the routing entries being cached, the throughput reached almost 4 PPC in a 4-width architecture. On the other hand, the queue size had little effect on the throughput improvement. A small queue with $Q = 16$ is enough for the 4-width architecture. 

\subsection{Overall Performance}

Based on the previous experiments, we estimate the overall performance of a 4-width 25-stage architecture. As Figure \ref{fig:ip_nos_rf4_largeleaf} shows, for the largest backbone routing table \textit{rrc11} with 154419 prefixes, each stage has fewer than 32K nodes. A 15-bit address is enough to index a node in the local memory of a stage. Since the pipeline depth is 25, we need an extra 5 bits to specify the distance. Thus, each node stored in the local memory needs 20 bits. The total memory needed for storing \textit{rrc11} in a 25-stage architecture is $20 \times 2^{15} \times 25 = 16$ Mb $= 2$ MB, where each stage needs 80 KB of memory. We use CACTI 4.2\cite{cacti} to estimate the memory access time and the power consumption. A 80 KB dual-port SRAM using 45 nm technology needs 0.53 ns to access, and dispatches 0.01 W of power. The maximum clock rate of the above architecture in ASIC implementation can be 1.87 GHz. Considering the throughput of 4 PPC as shown in Figure \ref{fig:ppc_iw}, the overall throughput can be as high as $4 \times 1.87$ = 7.5 G packets per second, i.e 2.4 Tbps for the minimum packet size of 40 bytes. Such a throughput is 14 times that of the state-of-the-art TCAM-based IP lookup engines \cite{ton06:zheng}. The overall power consumption is 0.25 W, which is only one eighth of that of the ``coolest'' TCAM solution \cite{infocom03:zane}. 

%


\section{Conclusions and Future Work}
\label{sec:conclusion}

This paper proposed a flexible dual-port SRAM based bidirectional linear pipeline architecture for scalable IP lookup and packet classification in IP routers. By using a bidirectional fine-grained mapping scheme, the tree nodes can be evenly distributed onto the pipeline stages. Due to its linear structure, the architecture can preserve the packet input order and support non-blocking route update. Using 2MB of memory to store a core routing table with over 150K entries, the architecture can sustain a high throughput of 0.6 Tbps and can further achieve 2.4 Tbps by employing caching.

For multi-dimensional packet classification, the operations in each stage are more complex than for the simple trie-based IP lookup. This may affect adversely the pipeline performance. We plan to develop new search data structures for packet classification so that pipelining can be more feasible. 

\bibliographystyle{abbrv}
\bibliography{biolp}  

\end{document}